\documentclass[a4paper,12pt]{article}

\usepackage{amsmath}
\usepackage{hyperref}
\usepackage{cite}
\usepackage{amssymb}
\usepackage{braket}
\title{Complementary ADHM Instanton Sigma Model}
\author{Abbas Ali\footnote{Email: aali.ph@amu.ac.in} ~and Mohsin Ilahi\\
		Physics Department, Aligarh Muslim University,\\ Aligarh-202002, India }
\date{}
\begin{document}

\maketitle

\begin{abstract}

We resolve the three decades old problem of the moduli space of ADHM instanton linear sigma model constructed by Witten in 1995. We construct an ADHM instanton linear sigma model that is complementary to Witten's original model in the sense of being dual to it. The resulting moduli space immediately suggests the solution to the mystery associated with the moduli space of the original model. Several problems are listed on which this new construction has  bearing.
\end{abstract}

More than three decades ago, Witten constructed a linear two-dimensional linear sigma model with a potential term and (0, 4) worldsheet supersymmetry in\cite{Witten:1994tz}. The interaction terms of this model satisfy the Atiyah-Drinfeld-Hitchin-Manin (ADHM) equations of self-dual Yang-Mills fields, that is, instantons\cite{Atiyah:1978ri}. This makes the former a candidate for stringy generalization of the latter. ADHM instantons carry a formidable reputation. Part of the reason for this reputation is that the ADHM construction is exceptionally brief and incredibly linear. Mathematicians continue their efforts to fathom its structure. In this regard we refer the reader to the following references  \cite{Rawnsley:1978mv, Lindenhovius:2011iac, Atanasov:2016iac, Donaldson:2022}. The aura of ADHM instantons being recondite is partly also due to the wrong impression that ADHM construction is a solution to Yang-Mills equations that are second-order coupled partial differential equations. This impression was corrected by Witten in Ref. \cite{Witten:1978qe} soon after the original ADHM paper. He pointed out that the ADHM construction is a solution to the self-dual Yang-Mills equations and these are first-order differential equations. On the physics side we again know a lot about the structure of the ADHM instantons see, for example, Refs.\cite{Christ:1978jy, Corrigan:1978ce,Corrigan:1983sv, Weinberg:2012pjx}. To our great relief these references clarify that the issue at hand basically concerns just the process of inverting a matrix equation involving matrices with quaternion entries.  

The ADHM instanton, just like the $SU(2)$ 't Hooft instanton, is a finite energy, as opposite to perturbatively small energy, classical solutions to field theory equations. Ever since their discovery in 1975 the 't Hooft instantons too have been studied very extensively. These were quantized in Ref.\cite{tHooft:1976snw} and their applications are wide ranging\cite{Coleman:1978ae}. These are amenable to a stringy generalization \cite{Strominger:1990et, Callan:1991at, Callan:1991dj, Callan:1991ky}. The original solution depends on $5k+4$ parameters. It is well known that the most general instanton solution should depend on $8k$ parameters for the $SU(2)$ gauge group. The ADHM construction is most general in the sense that in case of $SU(2)$ gauge group it depends on the full set of $8k$ parameters.

After corresponding stringy generalization was constructed by Witten in Ref.\cite{Witten:1994tz} its quantization was done by Lambert in Ref. \cite{Lambert:1995dp} (see also  Ref. \cite{Lambert:1995hs}). D-branes were included in the ambit in Refs.\cite{ Douglas:1996uz, Lambert:1996yd, Lambert:1997gs, Johnson:1998yw}. 

The moduli space of this model happens to be rather mysterious. This is captured in the potential for the construction for the one instanton case:
\begin{equation}
    V_W = \frac{1}{8}(X^{2}+\rho^{2})\phi^{2}.
    \label{potential1}
\end{equation}
The moduli space of this model is given by $\phi=0$ and any $X$. The mystery appears for the small instanton limit where $\rho\rightarrow 0$ and the potential for one instanton case becomes $V_W = \frac{1}{8}X^{2}\phi^{2}$. In this case one more branch of the moduli space appears for $X=0$ and any $\phi$ in addition to the original branch. 

In the original construction cause or the dynamics of the appearance of the additional branch was not clear. The problem was revisited  in Ref.\cite{Witten:1995zh} and a rationalization for the mysterious structure of the moduli space was presented. This rationalization should not be missed because of its hubris. From the analysis of the ADHM sigma model moduli space Witten concluded that the two branches are at a finite distance and these meet at $\phi=X=0$. But the issue has remained unsettled for nearly three decades.

In this note we present a resolution of the mystery of the moduli space of Witten's ADHM instanton linear sigma model. We present a complementary construction that was, in fact, suggested by Witten in his original paper. We then analyze the corresponding moduli space. We present a simple resolution of the three decades old mystery of the moduli space of Witten's ADHM instanton linear sigma model.

In this note we begin  by collecting the notation from original construction and Ref.\cite{Lambert:1995dp}. After discussing Witten's original ADHM instanton sigma model we discuss the construction of the  ADHM instanton sigma model that is complementary, in the sense to be discussed later, to the one constructed by Witten. The kinetic part of the action for the complementary model has the same form as that for the original model but the interaction part, including the potential, is different. We spell out in detail the difference in the Yukawa couplings of the original and the complementary models. 

We assert that the two models, Witten's original and our complementary models, are two independent branches of the ADHM instanton linear sigma models.

The two branches, the original and the complementary, are related by a disarmingly simple duality.  But this is a red herring in one sense. In particular we assert that given the original model the complementary one can not be worked out by a trained neural network, ML or artificial intelligence without human intervention.

In the process of working out of the details it becomes absolutely clear that the two branches of the ADHM instanton sigma model, the original and the complementary, are separate and independent of each other. This also explains the infinite distances between the two branches.

We then specialize, like Witten did in his construction, to one instanton case for $SU(2)$ group. After that we cover those aspects of the complementary model that are parallel with the original model. The comparison of the moduli space of the resulting model with the moduli space of the original model gives surprising and clear insight about the corresponding structure and above mentioned resolution of the mystery. We then summarize the additional results that we have obtained based on this experience. We conclude with the listing of our work under progress and future possibilities.

Let us begin with the action for Witten's original ADHM instanton linear sigma model.

 The action for ADHM sigma model is  
\begin{eqnarray}\label{action}
S &=& \int d^2 \sigma \biggl\{   \partial_{-}  {X}_{AY}\partial_{+} X^{AY}+ i\epsilon_{A'B'}\epsilon_{YZ}{\psi}_{-}^{A^{\prime}Y} \partial_{+} \psi_-^{B^{\prime}Z} \nonumber\\
&+&\partial_{-}{\phi}_{A^{\prime}Y^{\prime}}\partial_{+} \phi^{A^{\prime}Y^{\prime}}
+i\epsilon_{AB}\epsilon_{Y^{\prime}Z^{\prime}}\chi_-^{AY^{\prime}}\partial_{+} \chi_{-}^{BZ^{\prime}} 
+ i\lambda_+^a \partial_{-}\lambda_+^a \nonumber\\
&-& \frac{i}{2}m\lambda_+^a \Big ( \epsilon^{AB}\frac{\partial C^{a}_{AB'}}{\partial X^{BY}} \psi_-^{B'Y} 
+ \epsilon^{A'B'} \frac{\partial C^{a}_{AA'}}{\partial \phi^{B'Y'}} \chi_-^{AY'}\Big ) \nonumber \\
&-& \frac{1}{8}m^{2}\epsilon^{AB}\epsilon^{A'B'}C^a_{AA'}C^a_{BB'} \biggr\}
\end{eqnarray}
with 	
\begin{equation}  
\sigma^{\pm}=(\tau \pm \sigma)/2,~~~\partial_{\pm} = \frac{1}{\sqrt{2}}(\partial_0 \pm \partial_1)
\end{equation}
where $\tau$ and $\sigma$ are light-cone coordinates, the worldsheet metric being 
\begin{equation}
ds^2=d\tau^2-d\sigma^2
\end{equation}
with  $X^{AY}$, $A=1,2$ and $Y=1,2,\cdots,2k$ being $4k$ bosons are part of the standard multiplet and these have right handed superpartners $\psi_{-}^{A'Y}, A'=1,2$. We also have another set of $4k'$ bosons $\phi^{A'Y'}$, part of the twisted multiplet, with corresponding superpartners $\chi_{-}^{AY'}$ with $Y'=1,2,\cdots,2k'$. The raising and lowering of 
$Y, Z,\cdots$ and $Y', Z' ,\cdots$ indices is done with respective $Sp(k)$ and
$Sp(k')$ tensors $\epsilon^{YZ} (\epsilon_{YZ})$, $\epsilon^{Y'Z'} (\epsilon_{Y'Z'})$. Similarly the $A, B, \cdots$ and $A', B', \cdots$ indices  are raised (lowered) with the help of anti-symmetric tensors $\epsilon^{AB} (\epsilon_{AB})$, $\epsilon^{A'B'} (\epsilon_{A'B'})$ respectively. The free theory has an $F \times F' \times H \times H'$ symmetry that acts on $AB, A'B', YZ$ and $Y'Z'$ indices respectively where $F=SU(2)$, $F'=SU(2)'$, $H=Sp(k)$ and $H'=Sp(k')$. This is generally broken by the potential terms. 

To begin with there is a $Z_2$ symmetry between the standard and the twisted multiplet with respect to the interchange of $F$ and $F'$ (and $H$ and $H'$). In the original construction this was broken because only the   $F'$ symmetry was retained. As a result we got an action for $k'$ instanton number and $4k+4k'$ target space dimensions. This $Z_2$ symmetry will be broken in the present complementary construction too because we shall retain only th $F$ symmetry and get the action for $k$ instanton number and $4k+4k'$ target space dimensions.

At this point one issue has to be emphasized once again. The present complementary construction, though straightforward, is not trivial. In particular the task can not be assigned to a neural network nor we can use ML for this construction.

This chiral sigma model has $(0,4)$ on-shell supersymmetry
\begin{eqnarray}\label{susy}
\delta_{\eta} X^{AY}&=&i \epsilon_{A'B'}\eta^{AA'}_{+}\psi^{B'Y}_{-},
\delta_{\eta} \psi^{A'Y}_{-}= \epsilon_{AB}\eta^{AA'}_{+} \partial_{-} X^{BY}, \nonumber\\ 
\delta_{\eta} \phi^{A'Y'}&=&i \epsilon_{AB}\eta^{AA'}_{+}{\chi}^{BY'}_{-}, 
\delta_{\eta}{\chi}^{AY'}_{-} = \epsilon_{A'B'}\eta^{AA'}_{+} \partial_{-} \phi^{B'Y'}
\end{eqnarray}
with $\eta^{AA'}_{+}$ being an infinitesimal anti-commuting spinor parameter.
The reality conditions on the bosons are
\begin{equation}\label{boscond}
X^{AY}= \epsilon^{AB}\epsilon^{YZ} \overline {X}_{BZ}, ~~~~~~\phi^{A'Y'}= \epsilon^{A'B'}\epsilon^{Y'Z'} \overline {\phi}_{B'Z'}
\end{equation}
with similar conditions for $\psi_{-}^{A'Y}$ and $\chi_{-}^{AY'}$.

The supercharges $Q^{AA'}$ satisfy 
\begin{equation}\label{supercharges}
Q^{AA'}= \epsilon^{AB}\epsilon^{A'B'}{Q^{\dagger}_{BB'}}~~, ~~~~\lbrace Q^{AA'},Q^{BB'}\rbrace= \epsilon^{AB}\epsilon^{A'B'}P^{+}
\end{equation} 
with $P^{+}=P_{-}=-i\partial_{-}$. The supersymmetry transformations satisfy the algebra
\begin{equation}\label{susytrans}
[\delta_{\eta'},\delta_{\eta}]=-i \epsilon_{AB}\epsilon_{A'B'}\eta^{AA'}_{+} \eta{'}^{BB'}_{+} \partial_{-}.
\end{equation}

Before describing the complementary construction we shall make a digression to explain the supersymmetry structure involving the Yukawa coupling term in the original model. The Lagrangian for the left moving fermions is obtained in the following manner. One begins with the general Yukawa type interaction
\begin{equation}\label{yukint}
S_{\lambda}=-\frac{i}{2}\int d^{2}\sigma\lambda^{a}_{+}G_{a\theta}{\rho}^{\theta}_{-}
\end{equation}
with $\rho_{-}^{\theta}$ including all of $\psi's$ and $\chi's$. The corresponding equation of motion for $\lambda^{a}_{+}$ is
\begin{equation}\label{eom}
\partial_{-}\lambda^{a}_{+}=G^{a}_{\theta}{\rho}^{\theta}_{-}.
\end{equation}
The closure of the algebra (\ref{susytrans}) on $\lambda^{a}_{+}$ gives
\begin{equation}\label{cloalg}
[\delta_{\eta'},\delta_{\eta}]\lambda^{a}_{+}=-i \epsilon_{AB}\epsilon_{A'B'}\eta^{AA'}_{+} \eta{'}^{BB'}_{+} \partial_{-}\lambda^{a}_{+}.
\end{equation}
Using the equation of motion (\ref{eom}) the algebra (\ref{cloalg}) becomes
\begin{equation}\label{cloalgbec}
[\delta_{\eta'},\delta_{\eta}]\lambda^{a}_{+}=-i \epsilon_{AB}\epsilon_{A'B'}\eta^{AA'}_{+} \eta{'}^{BB'}_{+} \partial_{-}G^{a}_{\theta}{\rho}^{\theta}.
\end{equation}
Assuming a general supersymmetry transformation for the $\lambda^{a}_{+}$ one writes
\begin{equation}\label{susylamb}
\delta_{\eta}\lambda^{a}_{+}=\eta^{AA'}_{+}C^{a}_{AA'}
\end{equation}
with tensor $C^{a}_{AA'}$ being linear in $X's$ and $\phi's$ such that
\begin{equation}\label{deletaprime}
\delta_{\eta'}\delta_{\eta}\lambda^{a}_{+} 
= i\eta^{AA'}_{+}\biggl(\frac{\partial {C}^{a}_{AA'}}{\partial X^{BY}}\epsilon_{B'C'}\eta^{'BB'}_{+}\psi^{C'Y}_{-}
+\frac{\partial C^{a}_{AA'}}{\partial\phi^{B'Y'}}\epsilon_{BC}\eta^{'BB'}_{+}{\chi}^{CY'}_{-}\biggr).
\end{equation}
From (\ref{susylamb}) we also have
\begin{equation}
\delta_{\eta'}\lambda^{a}_{+}=\eta'^{BB'}_{+}C^{a}_{BB'}
\end{equation}
such that

\begin{equation}\label{deleta}
\delta_{\eta}\delta_{\eta'}\lambda^{a}_{+} 
= i\eta'^{BB'}_{+}\biggl(\frac{\partial {C}^{a}_{BB'}}{\partial X^{AY}}\epsilon_{A'C'}\eta^{AA'}_{+}\psi^{C'Y}_{-} +
\frac{\partial C^{a}_{BB'}}{\partial\phi^{A'Y'}}\epsilon_{AC}\eta^{AA'}_{+}{\chi}^{CY'}_{-}\biggr).
\end{equation}

Combining (\ref{deletaprime}) and (\ref{deleta}) one gets

\begin{eqnarray}\label{deletadeletap}
[\delta_{\eta'},\delta_{\eta}]\lambda^{a}_{+}&=&i \eta^{AA'}_{+} \eta{'}^{BB'}_{+}\biggl[ \left(\epsilon_{B'C'}\frac{\partial {C}^{a}_{AA'}}{\partial X^{BY}}
-\epsilon_{A'C'}\frac{\partial {C}^{a}_{BB'}}{\partial X^{AY}}\right)\psi^{C'Y}_{-}\nonumber\\&+&\left(\epsilon_{BC}\frac{\partial C^{a}_{AA'}}{\partial\phi^{B'Y'}}-\epsilon_{AC}\frac{\partial C^{a}_{BB'}}{\partial\phi^{A'Y'}}\right){\chi}^{CY'}_{-}\biggr].
\end{eqnarray}

The two terms on the RHS are independent and hence separately proportional to $\epsilon_{AB}\epsilon_{A'B'}$. These should vanish if we multiply these by $\delta^{AB}$ $\delta^{A'B'}$. This gives us the relations
\begin{eqnarray}\label{adhmcond1}
\frac{\partial C_{AA'}^a}{\partial X^{B Y}}
+\frac{\partial C_{BA'}^a}{\partial X^{A Y}}
=0=  \frac{\partial C_{AA'}^a}{\partial \phi^{B'Y'}}
+\frac{\partial C_{AB'}^a}{\partial \phi^{A'Y'}}.
\end{eqnarray}

The supersymmetry algebra above is satisfied if we take
\begin{equation}\label{gcond}
G^{a}_{\theta}\rho^{\theta} = \frac{1}{2}\biggl[\epsilon^{BD} \frac{\partial C^{a}_{BB'}}{\partial X^{DY}}\psi^{B'Y}_{-} + \epsilon^{B'D'}\frac{\partial C^{a}_{BB'}}{\partial \phi^{D'Y'}}{\chi}^{BY'}_{-}\biggr].
\end{equation}
This way one ends up with the Yukawa term in Eq.(\ref{action}) beginning with the ansatz (\ref{eom}).

The tensors $C^{a}_{AA'}$ obey
\begin{equation}\label{caaten}
\sum_{a}(C^{a}_{AA'}{C}^{a}_{BB'}+{C}^{a}_{BA'}{C}^{a}_{AB'})=0.
\end{equation}

To obtain the ADHM equations we begin the following general form for $	C^{a}_{AA'}$ that is linear in both $X$ and $\phi$
\begin{eqnarray}\label{gencaa}
C^{a}_{AA'}&=&M^{a}_{AA'}+\epsilon_{AB} N^{a}_{A'Y}X^{BY}+\epsilon_{A'B'}D^{a}_{AY'}\phi^{~B'Y'}\nonumber\\&+&\epsilon_{AB}\epsilon_{A'B'}E^{a}_{YY'}X^{BY}\phi^{~B'Y'}.
\end{eqnarray}

At this juncture we can bring in the construction for the complementary ADHM instanton linear sigma model.

To construct the complementary branch of the ADHM sigma model we shall maintain $F$ invariance and break $F'$ invariance by taking $M=D=0$ and thus we have
\begin{eqnarray}\label{mdzero}
\hat{C}^{a'}_{AA'}&=&\epsilon_{AB} N^{a'}_{A'Y}X^{BY}+ \epsilon_{AB}\epsilon_{A'B'}E^{a'}_{YY'}X^{BY}\phi^{~B'Y'}\nonumber\\&=&  X_{A}^{~Y}A^{a'}_{A'Y}(\phi)
\end{eqnarray}

where the tensor $A^{a'}_{A'Y}$ is linear in $\phi$. The condition in Eqn.(\ref{caaten}) this time becomes
\begin{equation}\label{aaten}
\sum_{a}(A^{a'}_{A'Y}(\phi)A^{a'}_{B'Z}(\phi)+A^{a'}_{B'Y}(\phi)A^{a'}_{A'Z}(\phi))=0.
\end{equation}
In the original construction $F'$ invariance was maintained and $F$ invariance broken by taking $M=N=0$ and with
\begin{eqnarray}\label{mnzero}
{C}^{a}_{AA'}&=&\epsilon_{A'B'}D^{a}_{AY'}\phi^{~B'Y'}+ \epsilon_{AB}\epsilon_{A'B'}E^{a}_{YY'}X^{BY}\phi^{~B'Y'}\nonumber\\&=& \phi_{A'}^{~Y'}B^{a}_{AY'}(X)
\end{eqnarray}
where $B^a_{AY'}(X)$ is linear in $X$ and we have 
\begin{equation}\label{bbcond}
\sum_{a}(B^{a}_{AY'}(X)B^{a}_{BZ'}(X)+B^{a}_{BY'}(X)B^{a}_{AZ'}(X))=0.
\end{equation}
There is a duality  between the two choices. This corresponds to interchange of $X \leftrightarrow \phi$, $ \psi \leftrightarrow \chi $, $k \leftrightarrow k'$ and $a \leftrightarrow a'$ \footnote{The complementary and original constructions can be termed A-and B-models respectively because of the terms $A^{a}_{A'Y}$ and $B^{a}_{AY'}$. This is not far fetched from the names used for topological A-and B-models obtained by twisting N=2 superconformal algebra.}. In a degenerate case the duality turns into a $Z_2$ symmetry. 

This is the sense in which the present model is complementary to Witten's original ADHM instanton linear sigma model. The two models are related by the above mentioned duality.

For the original construction the tensor $C^a$ is homogeneous and linear in $\phi$ and the potential quadratic. In particular the moduli space or the space of vacua is $\mathcal{M} = R^{4 k}$ ($X^{AY}, A=1,2$ and $Y=1,2,..,2k'$) that is given by $\phi=0$ and any $X$. For the complementary branch the corresponding tensor $\hat{C}^{a'}$ is linear and homogeneous in $X$ and the potential again quadratic. This time the moduli space is given by $X=0$ and any $\phi$ and thus it is $\mathcal{M'} = R^{4k'}$ ($\phi^{A'Y'},A'=1,2$ and $Y'=1,2,..,2k'$). Both $\mathcal{M}$  and $\mathcal{M'} $ have hyper-K\"ahler structure due to  (0, 4) supersymmetry. The $\phi \leftrightarrow X$ and part of the duality is apparent here. The $\psi \leftrightarrow \chi$ part will be apparent in a moment. 

On the original branch $\mathcal{M}$  of the moduli space, that is for $\phi=0$, the Yukawa couplings have the form $ \sum_{a}\lambda_{+}^aB^{a}_{AY'}\chi^{AY'}_-$ such that the fermionic partners of $X$, the $\psi_-$'s, are all massless. This is expected from  (0,4) supersymmetry.

On the complementary branch $\mathcal{M'}$ of the moduli space we have $X=0$ and the form of the Yukawa coupling is $ \sum_{a'}\hat\lambda_{+}^{a'}A^{a'}_{A'Y}\psi^{A'Y}_-$ and this time $\chi_-$, the fermionic partners of $\phi$, are all massless.

In the original model if the number $n$ of components of $\lambda_{+}^a$ is larger than $4k'$, the number of components of $\chi_{-}^{AY'}$, then generically all of the $\chi_-$'s get mass because of the Yukawa term and $N = n-4k'$ components of $\lambda_{+}$ are massless.  By supersymmetry all the $\phi$'s too become massive.

In the complementary model if the number $n$ of the left moving fermions $\hat\lambda^{a}_{+}$ is larger than $4k$, that is, the number of $\psi^{A'Y}_{-}$ ($A'=1,2, Y=1,2,...,4k$) then all components of $\psi_{-}$ get masses. By supersymmetry this happens when all components of $X$
get masses. The $N'=n'-4k$ components of $\hat\lambda^{a'}_{+}$ are massless.

We now work out the details for the complementary branch in parallel with the original construction. Let $\hat{v}^{a'}_{i'}$, $i=1,2,...,N'$ be a basis for massless components of $\hat\lambda^{a}_{+}$. These are solutions to the equation
\begin{equation}\label{mlclamb}
\sum_{a'}\hat{v}^{a'}_{i'}A^{a'}_{A'Y}=0.
\end{equation}
We shall choose $\hat{v}^{a'}_{i'}$ to be orthonormal, that is,
\begin{equation}\label{orthonorv}
\sum_{a'}\hat{v}^{a'}_{i'}\hat{v}^{a'}_{j'}=\delta_{i'j'}.
\end{equation}

The tensor $A^{a'}_{A'Y}(\phi)$ and hence $\hat{v}^{a'}_{i'}(\phi)$ are $\phi$-dependent. There is an $SO(N')$ transformation on the index $i'$. We shall take it  as a gauge invariance of the low energy theory. Setting
\begin{equation}\label{setlamb}
\hat\lambda^{a'}_{+}=\sum_{i=1}^{N'}\hat{v}^{a'}_{i'}\hat\lambda_{+i'}
\end{equation}
amounts to putting the massive modes of $\hat\lambda^{a'}_{+}$ to zero. Here $\hat\lambda_{+i'}$ are the massless left moving modes. Using $\partial_{-}(\hat{v}^{a'}_{j'}\hat\lambda_{+j'})=\partial_{-}\hat{v}^{a'}_{j'}\hat\lambda_{+j'}+\hat{v}^{a'}_{j'}\partial_{-}\hat\lambda_{+j'}$ and $\partial_{-}\hat{v}^{a'}_{j'}=\frac{\partial \hat{v}^{a'}_{j'}}{\partial \phi^{A'Y'}}\partial_{-}\phi^{A'Y'}$ the kinetic energy term for massless $\hat\lambda_{+i'}$ becomes
\begin{eqnarray}\label{kinmassless}
&& \frac{i}{2}\int d^{2}\sigma \sum_{i',j',a'}(\hat{v}^{a'}_{i'}\hat\lambda_{+i'})\partial_{-}(\hat{v}^{a'}_{j'}\hat\lambda_{+j})\nonumber\\ 
&=&   \frac{i}{2}\int d^{2}\sigma \sum_{i',j'}\lbrace\hat\lambda_{+i'}[\delta_{i'j'}\partial_{-} + A_{i'j'A'Y'}\partial_{-}\phi^{A'Y'}]\hat\lambda_{+j'}\rbrace
\end{eqnarray}
with 
\begin{equation}\label{stgaugef}
A_{i'j'A'Y'}=\sum_{a}\hat{v}^{a'}_{i'}\frac{\partial\hat{v}^{a'}_{j'}}{\partial \phi^{A'Y'}}
\end{equation}
which is the standard fermion coupling to spacetime gauge field $A_{i'j'A'Y'}$ (cf. Ref. \cite{Corrigan:1978ce}).

These gauge fields should be compatible with the hyper-K\"{a}hler structure on the moduli space $\mathcal{M}={R}^{4k'}$. For comparison the covariant derivative for the original branch was $\delta_{ij}\partial_{-}+ A_{ijAY}\partial_{-}X^{AY}$ and the expression for the instanton was
\begin{equation}\label{instanton}
A_{ijAY}=\sum_{a}v^{a}_{i}\frac{\partial v^{a}_{j}}{\partial X^{AY}}.
\end{equation}

The original construction was ADHM sigma model for an $SO(N)$ instanton with instanton number $k'$ given by the tensor $B^{a}_{AY'}(X)$ obeying Eqn.(\ref{bbcond}). It had a degeneracy implying  that components of $\phi$ are massive for all $X$. This will be apparent from the  expression for potential later on.

In the present construction we have got a complementary ADHM sigma model for an $SO(N')$ instanton with instanton number $k$ given by the tensor $A^{a}_{A'Y}(\phi)$ obeying Eqn.(\ref{aaten}). The degeneracy condition too is complementary and it says  that all the components of $X$ are massive for all $\phi$. The structure of the potential for the complementary branch makes it clear. 

Both the formulas, in Eqn. (\ref{stgaugef}) and Eqn. (\ref{instanton}), for ADHM instanton gauge fields are standard and complementary to each other in the sense that pervades the two constructions. Two constructions, Witten's original one and the one in this note, are dual to each other.

In the original model the two instantons are equivalent iff two $B^{a}_{AY'}$  tensors can be mapped onto each other by the action of $SO(n) \times Sp(k')$ that comes from the linear action of this group $a$ and $Y'$ indices of $\lambda^{a}_+$ and $\phi^{A'Y'}$ respectively. For the complementary construction two instantons are equivalent of the corresponding $A^{a'}_{A'Y}$ tensors are related by the transformation of the group  $SO(n') \times Sp(k)$ acting on the indices $a'$ and $Y$ of $\hat\lambda^{a'}_+$ and $X^{AY}$ respectively. We believe that the duality between the original and complementary branches is the same that becomes more apparent after Douglas' generalization to D-branes and that was pointed out in \cite{Lambert:1996yd,Lambert:1997gs}.

This is the construction of the complementary ADHM instanton sigma model in its generality. Now we shall specialize to the case for SU(2) group with $k=k'=1$. This is the complementary case or dual case to what Witten considered.  The $SU(2)$  instanton can be embedded in any larger group and the natural gauge groups in the two models are $SO(N')$ and $SO(N)$ respectively. We shall take $N'=N=4$ and use embedding of $SU(2)$  in $SO(4)$ given by $SO(4) = SU(2) \times SU(2)$. If we identify the one  instanton with one of the $SU(2)$'s then the other $SU(2)$ will be a trivial global symmetry group of the solution. The global $SU(2)$ for original and complementary constructions will be called $K$ and $K'$ respectively. We already have the global symmetry group $F'= SU(2)'$ for original and $F= SU(2)$ for complementary model.

Because of the rotation  groups $SO(4)$ of $R^4$  for the original model and $SO(4)'$ of $R^{'4}$ of complementary model we also have global groups $SO(4)= SU(2)_L \times SU(2)_R$ for orginal model and $SO(4)'=SU(2)^{'}_L \times SU(2)^{'}_R$  for the complementary model. Thus the total symmetry group for complementary  model is $F\times K' \times SU(2)'_L \times SU(2)'_R \cong SU(2)^4$  and for orginal model it is $F' \times K  \times SU(2)_L \times SU(2)_R \cong SU(2)^4$.  

For the $SU(2)$ instanton with intanton number there are four $X$'s and four $\phi$'s and a global symmetry group $F \times F' \times  H \times H'$ as identified in the beginning. In the original construction it was identified with $F' \times K  \times SU(2)_L \times SU(2)_R$. For the the complementary construction we shall identify it with $F \times K'  \times SU(2)'_L \times SU(2)'_R $.

To realize above global symmetry groups for the original model in the Lagrangian action of $F \times F' \times  H \times H'$ was needed on left moving fermions $\lambda^{a}_+$. To get  $N= 4$ with $N=n-4k'$ with $k'=1$ the needed value of $n$  was 8 and so is it with the complementary construction due to $N'=4$ and $N'=n-4k$ with $k=1$.

Experimentation lead to the realization that $\lambda_+$'s of the original model must be in $(\frac{1}{2},0,0,\frac{1}{2}) \oplus (0,0,\frac{1}{2},\frac{1}{2}) $. The $\lambda_{+}^{AY'}, A,Y=1,2$ and $\lambda_{+}^{YY'}, Y,Y'=1,2$ were chosen with the reality conditions  
\begin{eqnarray}\label{chorcond1}
\lambda^{AY'}_{+}= \epsilon^{AB}\epsilon^{Y'Z'} \overline {\lambda}_{+BZ'},~&&     \lambda^{YY'}_{+}= \epsilon^{YZ}\epsilon^{Y'Z'} \overline {\lambda}_{+ZZ'}.
\end{eqnarray} 

In complementary construction the suitable representation turns out to be $(0,\frac{1}{2},\frac{1}{2},0) \oplus (0,0,\frac{1}{2},\frac{1}{2})$ that is $\hat\lambda_{+}^{A'Y}$ and $\hat\lambda_{+}^{YY'}$ with reality conditions 
\begin{eqnarray}\label{chorcond2}
\hat\lambda^{A'Y}_{+}= \epsilon^{A'B'}\epsilon^{YZ} \overline {\hat\lambda}_{+B'Z},~&&     \hat\lambda^{Y'Y}_{+}= \epsilon^{YZ}\epsilon^{Y'Z'} \overline {\hat\lambda}_{+ZZ'}.
\end{eqnarray}

The Yukawa coupling tensor $\hat{C}^{a'}_{BB'}$ has two pieces 

\begin{eqnarray}\label{yukcten2}
\hat{C}^{Y'Y}_{~~~~BB'}= X_{B}^{~Y} \phi^{~Y'}_{B'},~ && 
\hat{C}^{A'Y}_{~~~~BB'}= \frac{\omega}{\sqrt{2}} \delta^{A'}_{~B'}X^{~Y}_{B}.  
\end{eqnarray}

Here $\omega$ is,  like $\rho$ of the original construction, the size of the complementary instanton. These expressions are complementary to the expressions for the original model.

In both the constructions the number of components of $\lambda_{+}$ and $\hat\lambda_+$ is $2 \times 2 + 2 \times 2=8$.  With the definitions of the original constructions

\begin{equation}\label{defocond}
X^{2}=\epsilon_{AB}\epsilon_{YZ}X^{AY}X^{BZ}, ~~  \phi^{2}=\epsilon_{A'B'}\epsilon_{Y'Z'}\phi^{A'Y'}\phi^{B'Z'}.
\end{equation}
The potentials for   our complementary complementary models becomes 
\begin{equation}\label{potential2}
 V_C = \frac{1}{8}(\phi^{2}+\omega^{2})X^{2}
\end{equation}
This should be compared with the potential for Witten's original model
given in Eqn. (\ref{potential1}).
In comparison the duality between the two constructions, $X\leftrightarrow \phi$, $\rho\leftrightarrow \omega$ is very clear.

To get the formula for instanton gauge field of the complementary construction in the familiar form the fermionic field ansatz is
\begin{eqnarray} \label{ferfa2}
\hat\lambda^{Y'Y} = \frac{\omega\hat\zeta^{~Y'Y}_+}{\sqrt{\omega^2 + \phi^2}},~ && \lambda^{A'Y} = -\frac{\sqrt{2}\phi^{A'}_{~~Y'}\hat\zeta_{+}^{~Y'Y}}{\sqrt{\omega^2 + \phi^2}}.
\end{eqnarray}
The kinetic energy for left moving fermions, then, becomes 
\begin{eqnarray} \label{kinlfer2}
&&\hat\lambda_{+Y'Y}\partial_{-}\hat\lambda^{Y'Y}_{+}+\hat\lambda_{+A'Y}\partial_{-}\hat\lambda^{A'Y}_{+} = \zeta_{+Y'Y}\partial_{-}\zeta^{~Y'Y}_{+} -\zeta_{+Y'Y}\times \nonumber\\&&\frac{\frac{1}{2}\epsilon_{A'B'}(\phi^{A'Y'}\partial_{-}\phi^{B'Z'}+\phi^{A'Z'}\partial_{-}\phi^{B'Y'})}{\phi^2+\omega^2}\hat\zeta^{Y}_{+Z'} 
\end{eqnarray}
which has the standard instanton expression. 

This concludes the discussion of the special case of one instanton for $SU(2)$ gauge group for the complementary model. As expected the expressions are complementary to the expressions for the original construction.

We shall now describe the main outcome of our construction by taking up the discussion of the moduli spaces. We shall talk about the moduli space of the original model, moduli space of the our complementary model and finally about the resolution of the mystery of the moduli space of the original model as pointed out by Witten in his paper on the original construction. We find that the nearly three decades old mystery has a very simple resolution.

Let us recall the structure of the moduli space of the original model. From the expression for the potential $V_W$ in  Eqn. (\ref{potential1}) we see that $\phi$ has an $X$ dependent mass term. We get the moduli space for $\phi = 0$ and any $X$. In the small instanton regime we have $\rho \rightarrow 0$ the potential becomes proportional to $\phi^2X^2$. In this case a new branch of the moduli space opens for $X = 0$ and any $\phi$. The two branches meet at $X=\phi=0$. This is the mystery of the moduli space of the original ADHM instanton sigma model constructed by Witten.

Witten revisited the problem of the moduli space in Ref. \cite{Witten:1995zh}.  Some new insights were gained but the mystery persisted. In particular by comparing with the Callan-Harvey-Strominger (CHS) instanton sigma model we realize that classically   the two branches of the moduli space are infinitely far away from each other.

Let us take up the discussion  of the moduli space of the complementary model. Initially it is parallel to the discussion for the original model. From the expression for the potential $V_C$ in  Eqn.(\ref{potential2}) we see that $X$ has a $\phi$ dependent mass term. Apart from that there is  moduli space for $X=0$ and any $\phi$. This is complementary to the structure in the original model, as expected. 

Even the supplementary analysis is in parallel with \cite{Witten:1995zh}. The metric obtained after including terms of order $\alpha'$ \cite{Strominger:1990et,Callan:1991dj}, for the complementary case, is given by 
\begin{equation}\label{instmet2}
ds^2=(d\phi)^2\Bigl(e^{2\phi_{0}}+8\alpha'\frac{\phi^2+2\omega^2}{(\phi^2+\omega^{2})^2} +... \Bigr).
\end{equation}

As in the original case the $X$ and $\phi$ branches are infinitely far away in the small instanton limit.

The moduli spaces of original and complementary models develop a second branch in the small instanton limit.
Thus we conclude that the mystery that was there in the moduli space Witten's original model is present in the complementary model too.

The metric in Eqn. (\ref{instmet2}) and its counter part for the original model develop a semi-infinite throat. This itself is a puzzle. This was already sorted out in Ref.\cite{Lambert:1995dp} where it was explained how to go from the ADHM moduli space to the CHS moduli space.

We now present the resolution to the nearly three decade old mystery of the moduli space of the original model. In the small instanton limit the the potentials of the two models coincide and the appearance of the new branch of the moduli space is the appearance of the moduli space of the other model. That is it is the moduli space of the complementary model that makes its appearance in the small instanton limit of the original model. Similarly the moduli space of Witten's original model makes an appearance in the small instanton limit of the complementary model. Thus the mystery of the moduli space of the ADHM instanton linear sigma model(s) has a very simple resolution.

Now the question of the infinite distance between two branches in the Callan-Harvey-Strominger analysis too has a natural explanation. It simply means that the two models are independent of each other. In case of the finite instanton there is a duality between the two models and that in particular means that the two models are independent of each other. In the small instanton limit this duality collapses to a $Z_2$ symmetry and hence the appearance of the second branch of the moduli space. This is consistent with the  well known fact that in the small instanton limit the theory gets an enhanced symmetry\cite{Witten:1995gx}.

We came to this problem to get some insight about the issues related to D1-D5 system and $AdS_3$ superstrings. In the $AdS_3/CFT_2$ version of the famous Maldacena conjecture, the AdS/CFT Correspondence \cite{Maldacena:1997re, Witten:1998qj, Gubser:1998bc}, the field theory dual to strings moving on $AdS_3 \times S^3 \times M^4$ with $M^4=K3$, $T^4$ and $S^3 \times S^1$ has small, middle and large N=4 superconfromal symmetry, respectively \cite{Sevrin:1988ab, Ivanov:1988rt, Ali:1993sd,  Ali:2000zu, Ali:2003aa,Ali:2000we, Hanany:2018hlz, Giveon:1998ns}. In this case  we have the long standing problem  of doubling of the Ramond superalgebra mentioned in the Appendix B of the Ref.\cite{Giveon:1998ns}. The complementary ADHM instanton sigma model constructed in this note suggests a way to solves this quarter of a century old problem.  The doubling of Ramond superalgebra in this case is due to the other  branch of the theory \cite{Ali:2023xov}. The two algebras are simply related to each other by the very same duality that we have uncovered here between Witten's original and our complementary ADHM instanton linear sigma model. The two branches are independent of each other. 

The same duality is at the core of the route to get to the proper $N=4$ superconformal free field theory that is relevant  for superstrings moving on $AdS_3 \times S^3 \times S^3 \times S^1$ backgrounds. In this case the duality becomes a $Z_2$ symmetry. In Ref. \cite{Gukov:2004ym} Gukov, Martinec, Moore and Strominger tried to obtain a complete free field realization of the large $N=4$ superconformal algebra but were stopped by a non-associativity.  In Ref.\cite{Ali:2023kkf} this two decades old problem has been resolved.

We have also done the quantization of the complementary model, in Ref.\cite{Lambert:1995dp},  along the lines adopted  for the quantization of the original ADHM instanton sigma model in Ref.\cite{Ali:2023zxt}.

It was also suggested in Witten's original paper to construct a model that contains both of the symmetries $F$ and $F'$. We have done that too in Ref. \cite{Ali:2023icn}. In Ref. \cite{Ali:2024amc} we have analyzed the symmetry structure of the ADHM instanton sigma models, $AdS_3$ superstrings and $N=4$ superconformal algebras.

We now conclude with a very brief discussion of future possibilities. We would also like to mention that it would be interesting to discuss  both of these sigma models for general groups as a  sigma model construction corresponding to \cite{Christ:1978jy} generalization of ADHM construction.  In \cite{Prabhakar:2023kfy}  Witten's ADHM instanton sigma model was formulated in projective superspace. The complementary model too can be cast in that formalism. In Ref. \cite{Ali:2025ntc} this has been formulated in the  harmonic superspace following Galperin-Sokatchev \cite{Galperin:1994qn}. Summary of has been presented in Ref. \cite{Ali:2025jcu}.

We would like to conclude with the following remark. It is well known that $N=4$ theories do not flow under renormalization group. The models considered in this note too have the same property \cite{Lambert:1995dp}. In view of that the journey from ADHM (or even 't Hooft) instanton to conformal field theory has a problem and we do not have any insight about it.

We also believe that the lines of investigation that have opened by the present construction will lead us to the solution to the problem of finding the holographic dual conformal field theory to superstrings  moving on $AdS_3 \times S^3 \times S^3 \times S^1$  backgrounds. This problem is as old as the AdS/CFT conjecture. In 2004 Gukov, Martinec, Moore and Strominger excluded all the proposals extent in this regard at that time. This and related problems have been analyzed in Refs. \cite{Hohenegger:2008du, Eberhardt:2017pty, Eberhardt:2017fsi, Dei:2018yth, Eberhardt:2019niq}). We believe that present investigations will shed light on all of these angles. This in turn should lead us to insights about holographic entanglement entropy in the context of $AdS_3/CFT_2$\cite{Jain:2017aqk, Jain:2017uhe, Malvimat:2018ood, Biswas:2023knx, Karch:2025fky}.
In Ref.\cite{Papadopoulos:2024uvi} Papdopoulos and Witten gave a direct proof of the fact that in two dimensions scale invariance implies conformal invariance. In Ref. \cite{Witten:2024yod} Witten investigated the moduli space of instantons on $S^3\times S^1$ and concluded that a large $N=4$ superconformal symmetry is relevant for superstrings moving in $AdS_3\times S^3\times S^3\times S^1$. These investigations are relevant for the issues discussed in the present and related notes. Of course $AdS_3$ superstrings are closely related to stringy generalization \cite{Ali:1992mj} of the BTZ blackhole \cite{Banados:1992wn} and the investigations along the lines advocated in this note will have implications for this aspect too. We intend to return to some of these investigations in future.

\textit{Acknowledgments}: This work was carried out as part of Mohsin Ilahi's Ph.D. thesis. The authors thank  P.P Abdul Salih and Shafeeq Rahman Thottoli for discussions.


\begin{thebibliography}{99}

\bibitem{Witten:1994tz}
E.~Witten,
J. Geom. Phys. \textbf{15}, 215-226 (1995)
[arXiv:hep-th/9410052 [hep-th]].
    
	\bibitem{Atiyah:1978ri}
	M.~F.~Atiyah, N.~J.~Hitchin, V.~G.~Drinfeld and Y.~I.~Manin,
	``Construction of Instantons,''
	Phys. Lett. A \textbf{65} (1978) 185-187.
 
\bibitem{Rawnsley:1978mv}
J.~H.~Rawnsley,
``Differential Geometry of Instantons", Lecture Notes, 1978.

\bibitem{Lindenhovius:2011iac}
A.~J.~Lindenhovius, ``Instantons and the ADHM Construction", M.Sc. Thesis [Advisor : R.Dijkgraaf], 2011.

\bibitem{Atanasov:2016iac}
A.~Atanasov, ``Instantons and ADHM Construction", Lecture Notes, 2016.

\bibitem{Donaldson:2022}S. Donaldson, ``The ADHM construction of Yang-Mills instantons",  [arXiv:2205.08639v1 [math.DG]].

	\bibitem{Witten:1978qe}
	E.~Witten,
	``Some Comments on the Recent Twistor Space Constructions,'' Workshop on Applications of Complex Manifold Techniques to Problems in Theoretical Physics, 1978.


	
	\bibitem{Christ:1978jy}
	N.~H.~Christ, E.~J.~Weinberg and N.~K.~Stanton,
	``General self-dual Yang-Mills Solutions,''
	Phys. Rev. D \textbf{18} (1978) 2013.

\bibitem{Corrigan:1978ce}
E.~Corrigan, D.~B.~Fairlie, S.~Templeton and P.~Goddard,
``A Green's Function for the General Self-dual Gauge Field,''
Nucl. Phys. B \textbf{140} (1978) 31-44.


	\bibitem{Corrigan:1983sv}
	E.~Corrigan and P.~Goddard,
	``Construction of Instanton and Monopole Solutions and Reciprocity,''
	Annals Phys. \textbf{154} (1984) 253.

\bibitem{Weinberg:2012pjx}
E.~J.~Weinberg,
``Classical solutions in quantum field theory: Solitons and Instantons in High Energy Physics,''
Cambridge University Press, 2012.
 
\bibitem{tHooft:1976snw}
G.~'t Hooft,
``Computation of the Quantum Effects Due to a Four-Dimensional Pseudoparticle,''
Phys. Rev. D \textbf{14} (1976) 3432-3450,
[erratum: Phys. Rev. D \textbf{18} (1978) 2199].

\bibitem{Coleman:1978ae}
S.~R.~Coleman,
``The Uses of Instantons,''
Subnucl. Ser. \textbf{15} (1979) 805,
HUTP-78-A004.

\bibitem{Strominger:1990et}A. Strominger,``Heterotic Solitons'', Nucl. Phys. B \textbf{343} (1990) 167.

	\bibitem{Callan:1991dj}
	C.G. Callan, J.A. Harvey and A. Strominger,  ``Worldsheet Approach to Heterotic Instantons and Solitons'',
	Nucl. Phys. B \textbf{359} (1991) 611.

\bibitem{Callan:1991ky}
C.~G.~Callan, Jr., J.~A.~Harvey and A.~Strominger,
``Worldbrane actions for string solitons,''
Nucl. Phys. B \textbf{367} (1991) 60-82.
 
	\bibitem{Callan:1991at}
	C.~G.~Callan, Jr., J.~A.~Harvey and A.~Strominger,
	``Supersymmetric string solitons,'' in: Proc. String Theory and Quantum Gravity '91(Trieste, 1991)
	[arXiv:hep-th/9112030].
 
\bibitem{Lambert:1995dp}
	N.~D.~Lambert,
	``Quantizing the (0,4) Supersymmetric ADHM Sigma Model,''
	Nucl. Phys. B \textbf{460} (1996) 221-232,
	[arXiv:hep-th/9508039].

\bibitem{Lambert:1995hs}
N.~D.~Lambert,
``Two loop renormalization of massive (p, q) supersymmetric sigma models,''
Nucl. Phys. B \textbf{469} (1996), 68-92
[arXiv:hep-th/9510130 [hep-th]].

	
\bibitem{Douglas:1996uz}
M.~R.~Douglas,
``Gauge fields and D-branes,''
J. Geom. Phys. \textbf{28} (1998) 255-262,
[arXiv:hep-th/9604198 [hep-th]].

	\bibitem{Lambert:1996yd}
	N.~D.~Lambert,
	``Heterotic p-branes from massive sigma models,''
	Nucl. Phys. B \textbf{477} (1996) 141-154, 
	[arXiv:hep-th/9605010].

 
	\bibitem{Lambert:1997gs}
	N.~D.~Lambert,
	``D-brane bound states and the generalized ADHM construction,''
	Nucl. Phys. B \textbf{519} (1998) 214-224, 
	[arXiv:hep-th/9707156].

\bibitem{Johnson:1998yw}
C.~V.~Johnson,
``On the (0,4) conformal field theory of the throat,''
Mod. Phys. Lett. A \textbf{13} (1998) 2463-2474,
[arXiv:hep-th/9804201 [hep-th]].

\bibitem{Witten:1995zh} E. Witten, ``Some Comments on String Dynamics" [hep-th/9507121], in Strings '95, Future Perspectives in String Theory, ed. I. Bars et. al.


	\bibitem{Witten:1995gx}
	E.~Witten,
	``Small instantons in string theory,''
	Nucl. Phys. B \textbf{460} (1996) 541-559,
	[arXiv:hep-th/9511030].

\bibitem{Maldacena:1997re}
J.~M.~Maldacena,
``The Large N limit of superconformal field theories and supergravity,''
Adv. Theor. Math. Phys. \textbf{2}, 231-252 (1998)
[arXiv:hep-th/9711200 [hep-th]].

\bibitem{Witten:1998qj}
E.~Witten,
``Anti-de Sitter space and holography,''
Adv. Theor. Math. Phys. \textbf{2}, 253-291 (1998)
[arXiv:hep-th/9802150 [hep-th]].

\bibitem{Gubser:1998bc}
S.~S.~Gubser, I.~R.~Klebanov and A.~M.~Polyakov,
``Gauge theory correlators from noncritical string theory,''
Phys. Lett. B \textbf{428}, 105-114 (1998)
[arXiv:hep-th/9802109 [hep-th]].

 
\bibitem{Sevrin:1988ab}
A. Sevrin, W. Troost and A. van Proeyen, ``Superconformal Algebras in Two-Dimensions with N=4'', Phys. Lett. B 208 (1988) 447.
		
		
\bibitem{Ivanov:1988rt}
E.~A.~Ivanov, S.~O.~Krivonos and V.~M.~Leviant,
``Quantum N=3, N=4 Superconformal WZW Sigma Models,''
Phys.\ Lett.\ B {\bf 215} (1988) 689,
Erratum: [Phys.\ Lett.\ B {\bf 221} (1989) 432].

\bibitem{Ali:1993sd}
A.~Ali and A.~Kumar,
Mod. Phys. Lett. A \textbf{8}, 1527-1532 (1993)
doi:10.1142/S0217732393001252
[arXiv:hep-th/9301010 [hep-th]].

\bibitem{Ali:2000zu}
A.~Ali,
``Free Field Realizations of N=4 Superconformal Algebras,''
Indian J. Pure Appl. Phys. \textbf{38} (2000) 446-452.
			
\bibitem{Ali:2003aa}
A.~Ali,
``Types of Two-dimensional N = 4 Superconformal Field Theories,''
Pramana \textbf{61} (2003) 1065-1078,
[arXiv:hep-th/9906096].

			
\bibitem{Ali:2000we}
A.~Ali, ``Conformal Symmetry of Superstrings on $AdS_3 \times S^3 \times T^4$ and D1 / D5 system,''
Mod.\ Phys.\ Lett.\ A {\bf 17} (2002) 2477,
[hep-th/0007021].
		
			
\bibitem{Hanany:2018hlz}
A.~Hanany and T.~Okazaki,
``(0,4) brane box models,''
JHEP \textbf{03} (2019) 027,
[arXiv:1811.09117 [hep-th]].

   
		
\bibitem{Giveon:1998ns}
	A.~Giveon, D.~Kutasov and N.~Seiberg,
	``Comments on String Theory on $AdS_3$,''
	Adv.\ Theor.\ Math.\ Phys.\  {\bf 2} (1998) 733,
	[hep-th/9806194].

\bibitem{Ali:2023xov}
A.~Ali,
``Superalgebra Doubling in $AdS_3$ Superstrings,''
[arXiv:2306.11047 [hep-th]].

\bibitem{Gukov:2004ym}
S.~Gukov, E.~Martinec, G.~W.~Moore and A.~Strominger,
``The Search for a holographic dual to $AdS_3\times S^3\times S^3\times S^1$,''
Adv. Theor. Math. Phys. \textbf{9} (2005), 435-525
[arXiv:hep-th/0403090 [hep-th]].

\bibitem{Ali:2023kkf}
A.~Ali and M.~Ilahi,
``$Z_2$ Symmetry of $AdS_3 \times S^3 \times S^3 \times S^1$ Superstrings,''
[arXiv:2306.13970 [hep-th]].



\bibitem{Ali:2023zxt}
A.~Ali, M.~Ilahi and S.~R.~Thottoli,
``Quantization of Complementary ADHM Sigma Model,''
[arXiv:2306.10002 [hep-th]].

\bibitem{Ali:2023icn}
A.~Ali and P.~P.~A.~Salih,
``Complete ADHM Sigma Model,''
[arXiv:2305.09516 [hep-th]].

\bibitem{Ali:2024amc}
A.~Ali, M.~Ilahi, P.~P.~A.~Salih and S.~R.~Thottoli,
``Symmetry Structure of ADHM Sigma Models and $AdS_3$ Superstrings,''
[arXiv:2409.00474 [hep-th]].

\bibitem{Prabhakar:2023kfy}
N.~S.~Prabhakar and M.~Ro\v{c}ek,
``(0, 4) Projective superspaces. Part I. Interacting linear sigma models,''
JHEP \textbf{07} (2023) 117,
[arXiv:2303.14675 [hep-th]].

\bibitem{Ali:2025ntc}
A.~Ali, M.~Ilahi, P.~P.~A.~Salih and S.~R.~Thottoli,
``Harmonic Superspace for Ali-Ilahi's ADHM Instanton Sigma Model,''
[arXiv:2507.22948 [hep-th]].

\bibitem{Galperin:1994qn}
A.~Galperin and E.~Sokatchev,
``Manifest supersymmetry and the ADHM construction of instantons,''
Nucl. Phys. B \textbf{452} (1995) 431-455,
[arXiv:hep-th/9412032].

\bibitem{Ali:2025jcu}
A.~Ali, M.~Ilahi, P.~P.~A.~Salih and S.~R.~Thottoli,
``Off-shell Formalism for Ali-Ilahi's ADHM Instanton Sigma Model,''
[arXiv:2507.11305 [hep-th]].


\bibitem{Hohenegger:2008du}
S.~Hohenegger, C.~A.~Keller and I.~Kirsch,
``Heterotic AdS(3) / CFT(2) duality with (0,4) spacetime supersymmetry,''
Nucl. Phys. B \textbf{804}, 193-222 (2008)
[arXiv:0804.4066 [hep-th]].

\bibitem{Eberhardt:2017pty}
L.~Eberhardt, M.~R.~Gaberdiel and W.~Li,
``A holographic dual for string theory on AdS$_{3}$\texttimes{}S$^{3}$\texttimes{}S$^{3}$\texttimes{}S$^{1}$,''
JHEP \textbf{08}, 111 (2017)
[arXiv:1707.02705 [hep-th]].

\bibitem{Eberhardt:2017fsi}
L.~Eberhardt, M.~R.~Gaberdiel, R.~Gopakumar and W.~Li,
``BPS spectrum on AdS$_3\times $S$^3 \times $S$^3 \times $S$^1$,''
JHEP \textbf{03}, 124 (2017)
[arXiv:1701.03552 [hep-th]].

\bibitem{Dei:2018yth}
A.~Dei, M.~R.~Gaberdiel and A.~Sfondrini,
``The plane-wave limit of ${\rm AdS}_3 \times {\rm S}^3 \times {\rm S}^3 \times {\rm S}^1$,''
JHEP \textbf{08}, 097 (2018)
[arXiv:1805.09154 [hep-th]].

\bibitem{Eberhardt:2019niq}
L.~Eberhardt and M.~R.~Gaberdiel,
``Strings on $\text{AdS}_3 \times \text{S}^3 \times \text{S}^3 \times \text{S}^1$,''
JHEP \textbf{06}, 035 (2019)
[arXiv:1904.01585 [hep-th]].

\bibitem{Jain:2017aqk}
P.~Jain, V.~Malvimat, S.~Mondal and G.~Sengupta,
``Holographic entanglement negativity conjecture for adjacent intervals in $AdS_3/CFT_2$,''
Phys. Lett. B \textbf{793}, 104-109 (2019)
[arXiv:1707.08293 [hep-th]].

\bibitem{Jain:2017uhe}
P.~Jain, V.~Malvimat, S.~Mondal and G.~Sengupta,
``Covariant holographic entanglement negativity for adjacent subsystems in AdS$_3$ /CFT$_2$,''
Nucl. Phys. B \textbf{945}, 114683 (2019)
[arXiv:1710.06138 [hep-th]].

\bibitem{Malvimat:2018ood}
V.~Malvimat, S.~Mondal, B.~Paul and G.~Sengupta,
``Covariant holographic entanglement negativity for disjoint intervals in $AdS_3/CFT_2$,''
Eur. Phys. J. C \textbf{79}, no.6, 514 (2019)
[arXiv:1812.03117 [hep-th]].

\bibitem{Biswas:2023knx}
S.~Biswas, A.~Dey, B.~Paul and G.~Sengupta,
``Covariant odd entanglement entropy in AdS$_3$/CFT$_2$,''
[arXiv:2312.12829 [hep-th]].

\bibitem{Karch:2025fky}
A.~Karch, H.~Ooguri and M.~Wang,
``Nonrenormalization Theorem for ${\cal N}=(4,4)$ Interface Entropy,''
[arXiv:2502.06928 [hep-th]].

\bibitem{Papadopoulos:2024uvi}
G.~Papadopoulos and E.~Witten,
``Scale and conformal invariance in 2d \ensuremath{\sigma}-models, with an application to $\mathcal{N}$ = 4 supersymmetry,''
JHEP \textbf{03}, 056 (2025)
[arXiv:2404.19526 [hep-th]].

\bibitem{Witten:2024yod}
E.~Witten,
``Instantons and the large ${\mathcal{N}} = 4$ algebra,''
J. Phys. A \textbf{58}, no.3, 035403 (2025)
[arXiv:2407.20964 [hep-th]].

\bibitem{Ali:1992mj}
A.~Ali and A.~Kumar,
``O (d, d) transformations and 3-D black hole,''
Mod. Phys. Lett. A \textbf{8}, 2045-2052 (1993)
[arXiv:hep-th/9303032 [hep-th]].

\bibitem{Banados:1992wn}
M.~Banados, C.~Teitelboim and J.~Zanelli,
``The Black hole in three-dimensional space-time,''
Phys. Rev. Lett. \textbf{69}, 1849-1851 (1992)
[arXiv:hep-th/9204099 [hep-th]].

\end{thebibliography}
\end{document}